# Cellular forgetting, desensitisation, stress and aging in signalling networks. When do cells refuse to learn more?

Tamás Veres[1,†], Márk Kerestély[1,†], Borbála M. Kovács[1], Dávid Keresztes[1], Klára Schulc[1,2], Erik Seitz[1], Zsolt Vassy[1], Dániel V. Veres[1,3], Peter Csermely[1*]

**Abstract**
Recent findings show that single, non-neuronal cells are also able to learn signalling responses developing cellular memory. In cellular learning nodes of signalling networks strengthen their interactions e.g. by the conformational memory of intrinsically disordered proteins, protein translocation, miRNAs, lncRNAs, chromatin memory and signalling cascades. This can be described by a generalized, unicellular Hebbian learning process, where those signalling connections, which participate in learning, become stronger. Here we review those scenarios, where cellular signalling is not only repeated in a few times (when learning occurs), but becomes too frequent, too large, or too complex and overloads the cell. This leads to desensitisation of signalling networks by decoupling signalling components, receptor internalization, and consequent downregulation. These molecular processes are examples of anti-Hebbian learning and 'forgetting' of signalling networks. Stress can be perceived as signalling overload inducing the desensitisation of signalling pathways. Aging occurs by the summative effects of cumulative stress downregulating signalling. We propose that cellular learning desensitisation, stress and aging may be placed along the same axis of more and more intensive (prolonged or repeated) signalling. We discuss how cells might discriminate between repeated and unexpected signals, and highlight the Hebbian and anti-Hebbian mechanisms behind the fold-change detection in the NF-$\kappa$B signalling pathway. We list drug design methods using Hebbian learning (such as chemically-induced proximity) and clinical treatment modalities inducing (cancer, drug allergies) desensitisation or avoiding drug-induced desensitisation. A better discrimination between cellular learning, desensitisation and stress may open novel directions in drug design, e.g., helping to overcome drug-resistance.

**Keywords** allergy, asthma, diabetes, habituation, heat shock, heart failure, incoherent type-1 feedforward loop; metabolon, mnemon, prion, protein translocation; receptor downregulation, scaffold proteins

**Learning of signalling networks – at the level of their components**

Molecular mechanisms of neuronal learning became well established [1]. However, much less is known about the regulation of learning at the individual, non-neuronal cells. Recent findings gave further evidence that learning, indeed occurs in unicellular organisms, as well as in individual cells of various tissues other than neurons, even in rather sophisticated forms [2]. In our paper we define cellular learning as an adaptive response to a stimulus, when the stimulus is repeated in a short time. This leaves out many classical models of learning (such as Pavlovian conditional learning) from our discussion. However, such a simplification greatly helps the identification of molecular mechanisms, which become increasingly obscured when long-term, multistep adaptation phenomena are examined, such as cell differentiation or tumour development. Several experiments in budding yeast, *Arabidopsis* or rice cells, mouse fibroblasts or murine CD8$^+$ memory cells showed the formation of molecular memory resulting in a faster, larger, more sensitive and/or more robust response after the second signal than the first [3-9].

---

[*] Peter Csermely, csermely.peter@med.semmelweis-univ.hu
[†] These authors contributed equally to the paper: Tamás Veres, Márk Kerestély
[1] Department of Molecular Biology, Semmelweis University, Budapest, Hungary
[2] Division of Oncology, Department of Internal Medicine and Oncology, Semmelweis University, Budapest, Hungary
[3] Turbine Ltd, Budapest, Hungary

Various molecular mechanisms induce a faster and stronger response after a repeated signal in single cells. We mention the conformational memory of intrinsically disordered proteins (IDPs) first, where the IDP transiently keeps its ordered conformation acquired after the first signal, and if the second signal arrives within the time window of the IDPs relaxation back to the disordered state, than the second signal finds the IDP in a 'conformationally-primed', 'memory'-state [10,11]. IDPs may act like molecular switches changing the direction of signal transmission [12]. Prions are an important class of IDPs. Prion proteins may transform themselves to a β-sheet enriched prion form, which forms aggregates. In budding yeast cells the prion form of Pin1 maintained the molecular memory of a previous heat shock for many generations [4]. Oligomerizing proteins involved in cellular memory formation were called as mnemons [13,14]. In case of the Whi3 protein present in yeast mnemon and prion states were shown to be associated, which confines the memory of deceptive courtship to the mother cell [15]. Increased association of 'conformationally-primed' IDPs with their signalling partners can be regarded as an increased network edge weight of signalling networks [10,16].

Signal-induced protein translocation (e.g. between the cytoplasm and the mitochondria or cell nucleus) is a widespread phenomenon in the cell potentially involving thousands of human proteins [17]. Nuclear residence time of the yeast cyclin Cln3 finely tunes Whi5 inactivation by phosphorylation. Whi5 is re-activated rapidly with a half-time of ~12 min. Thus the Cln3/Whi5 system provides a rapidly changing short term memory of environmental nutrient levels for yeast cells [18]. Mitochondrial translocation of the adaptor protein p66SHC was associated with the formation of hyperglycaemic cellular memory of human aortic endothelial cells [19]. Conversely, inhibition of nuclear translocation of NK-κB p65 disrupted the formation of CD8$^+$ memory T and memory B cells [20,21]. Translocating proteins build a number of new connections in signalling networks, which, again, shows a large increase of all signalling network edge weights involved.

MicroRNAs are involved both in sensitisation- and habituation-type cellular memory formation. MiRNA-156 participated in the molecular memory formation of previous heat shock in *Arabidopsis* cells lasting for several days [6]. As an additional example for sensitisation, miRNA-21 preserved the fibrotic mechanical memory of mesenchymal stem cells [22]. As an example of habituation, both miRNA-221 and miRNA-222 were involved in the memory development of lipopolysaccharide tolerance [23]. Long non-coding RNAs (lncRNAs) played an important role in the memory formation of rice cells after drought stress [7] and the formation of CD8$^+$ memory T cells after lymphocytic choriomeningitis virus infection [24]. Increased miRNA and lncRNA levels correspond to increased network edge weights of miRNA connections in signalling networks.

**Learning of signalling networks – at the network level**

After the contribution of single macromolecules (proteins and RNAs) to the formation of cellular memory, we give additional three examples of more complex, system-level, signalling network-type adaptation. The first example is that of epigenetic mechanisms and chromatin memory [25]. A large variety of histone modifications and DNA methylation constitute transcriptional memory. Histone H3 lysine methylation was shown to participate in the cellular memory development of yeast [26], *Arabidopsis* [5] (specifically mediated by heat shock factors HSFA2 and HSFA3; 27], mouse fibroblast and HeLa cell sensitisation to IFN-β and -γ, respectively [8,28], as well as in CD8$^+$ memory T cell formation [9]. DNA methylation pattern of 132 genes were changed in the development of CD4$^+$ memory T cells [29]. CRISPRoff, a single dead Cas9 fusion protein establishes DNA methylation and repressive histone modifications providing a genome-wide transcriptional memory [30]. Several studies showed the involvement of three-dimensional chromatin structure reorganization during cellular memory development in yeast [3,31], as well as in the sensitisation to repeated IFN-γ treatment of HeLa cells [28] by the same, nuclear pore protein 100/98-mediated chromatin-reorganization process [32].

The second example of network-type cellular memory formation is that of signalling protein kinase cascades. Members of the Hog1 signalling pathway of osmotic stress in yeast remained phosphorylated even after minutes of the first stress, and 'waited' pre-activated for a faster response to a potential repeated stress [33]. Similarly, in the mitogen-activated protein kinase (MAPK) cascade different relaxation rates of individual components developed an 'activation-competent' state inducing post-activation protein phosphorylation bursts [34]. MAPK pathway members are organized by pathway scaffold proteins from fungi (e.g. Far1- and Ste5-like proteins [35]) to humans (e.g. RACK1 [36]). These proteins, once they became activated, maintain larger pathway segments pre-organized, ready to respond to the second stimulus faster, and stronger. We note that similar signalling cascade memories may be postulated in each signalling pathway. As examples the JNK and Hippo pathway cascades are enhanced by the scaffolding proteins JIP1 and MOB1A, respectively [37,38]. These scaffolds may prime these pathways giving a stronger second response after an initial stimulus.

The third example expands the above idea of pathway organization and consequent cellular memory formation to networks other than signalling networks, such as metabolic networks. Analysis of non-Markovian chemical reaction networks on gene expression showed that molecular memory of protein synthesis and degradation may induce feedback, bimodality and switch behaviour, and may fine tune gene expression noise, all components of molecular memory [39]. Even bacteria use their inner membrane as a scaffold [40], as well as bacterial microcompartments [41] to enhance the metabolic flux of their enzymes. Mitochondria and other intracellular compartments also function as eukaryotic organizers of metabolic processes [42]. Metabolons are multienzyme complexes that are held together by noncovalent interactions enhancing their cooperation and summative metabolic flux by substrate channelling in e.g. glycolysis, branched chain amino acid oxidation, purine biosynthesis, etc. [43–47]. All these bacterial, mitochondrial and cellular microcompartmental metabolic scaffolds, as well as metabolons are potential organizers of cellular memory.

**Hebbian learning of signalling networks**

Practically all molecular mechanisms of cellular memory formation mentioned above are satisfying the basic concept of Hebbian learning, i.e.: the increase of the connection strength of those learning components (in the initial concept: neurons) which are involved in the learning process [1,48]. Stronger and faster binding of 'conformationally primed' IDPs, prions and mnemons, protein translocation, overexpression of miRNAs and lncRNAs, chromatin memory, scaffolded protein kinase and metabolic pathways are all examples of connection strength increases after an initial signal – serving as potential learning mechanisms of non-neuronal single cells (Fig 1).

Obviously, single cells can not express the complexity of the learning process of multicellular networks. This is especially true to that of neuronal networks. One simple reason of this is the number of connections. While macromolecules may bind only handful of other macromolecules, neurons developed axons and dendrites, which (by their tremendously increased surface areas) allow their connections to tens of thousands of other neurons. This by itself already magnifies the structural complexity which may be achieved by neuronal networks, and allows the development of incomparably more sophisticated learning processes than those of single, non-neuronal cells.

**Anti-Hebbian learning in non-neuronal cells**

Hebbian learning needs to be complemented by the reverse process, where connection strengths decrease, since only Hebbian-type 'positive' changes of the system would lead to the system's over-excitation. This was first generally formalized by Oja's rule, which keeps the total of connection strengths constant during a Hebbian learning process [49]. In cells connection strength decrease (decay) is generally introduced by cellular noise [50]. However, anti-Hebbian learning may also decrease the strength of specific connections, such as the reduced expression of the *STL1* sugar transporter gene in budding yeast cells after hyperosmotic stress [31], the diminished response of *MYC*-dependent genes after repeated dehydration stress in *Arabidopsis* [51] and the immune tolerance of macrophages after repeated lipopolysaccharide exposure [23]. Several of these direct, anti-Hebbian molecular mechanisms lead to habituation, where the cell displays a decreased response to repeated stimulation. Biological pathway network models were able to display both sensitisation (Hebbian) and habituation (anti-Hebbian) behaviour [52].

**Cellular memory and forgetting**

As first suggested by François Jacob and Jacques Monod in 1961 [53], individual cells also have a memory, i.e. the persistence of a cellular state, which is acquired after a stimulus. Cellular memory allows the establishment and maintenance of the identity of individual cells in heterogeneous cellular populations. Memory of the cells is manifested by bistable feedback loops or epigenetic marks conferring hysteresis and simple cognitive functions to cellular behaviour [25,54,55]. Intertwined feedback loops reduce cellular noise [50] and induce hysteresis, since stabilization of the 'signalling-on' state creates a resistance to return to the initial, 'signalling-off' state. While simple, hysteresis-type memory can be maintained by self-sustaining feedback loops, there seems to be a minimal network size requirement of at least 5 nodes to display richer memory functions [52].

Forgetting of organisms such as *C. elegans* is induced by cellular mechanisms, like the Musashi, MSI1-induced down-regulation of the ARP2/3 complex (playing a major role in the organization of the cytoskeleton) [56]. Increased cellular noise is a key factor of 'forgetting' in single, non-neuronal cells [50]. As a more specific example for the effect of cellular noise, robustness of the MAPK pathway becomes reduced, if environmental fluctuations (extrinsic noise) or variances of inherent chemical reaction rates (intrinsic noise) grow beyond a certain threshold [57].

We list three examples, where individual molecular mechanisms are involved in cellular 'forgetting'. First, erasure of DNA-methylation is meditated by ten-eleven translocation (TET) DNA-demethylases [58]. Second, the long noncoding RNA, originating at –2700 upstream of the budding yeast HO endonuclease, erased previous molecular memory of nutrient deprivation- or pheromone-induced cell cycle arrest [59]. Third, molecular chaperones may help the disorganization of protein sequences, thus they may act as facilitators of both molecular memory formation and cellular 'forgetting' [60]. However, currently beyond these mechanisms we do not know enough about the molecular systems regulating 'forgetting' in individual, non-neuronal cells.

We note that anti-Hebbian learning diminishes the strength of certain molecular connections, while cellular 'forgetting' may also induce a more general decrease of connection strengths. However, there is an obvious overlap between the two phenomena.

**Desensitisation of signalling networks**

Desensitisation of signalling responses is a general, habituation-type regulatory mechanism of signalling pathways. The most widespread way of desensitisation is receptor down-regulation by internalization (many times involving autophagy) and consequent degradation. We list here only a few examples of the many: the key plant stress signalling hormone, abscisic acid is desensitised by numerous steps of directed protein degradation [61]. An early example was the desensitisation of protein kinase C by its nonmetabolizable, long-term agonist, phorbol ester [62].

As an archetype of desensitisation G-protein-coupled receptor (GPCR) kinases (GRKs) induce arrestin binding to GPCRs, dissociating G-proteins and leading to GPCR internalization [63,64]. While short term activation of GPCRs causes receptor desensitisation *via* β-arrestin-mediated decoupling from G proteins, long-term (hours to days) activation induces receptor down-regulation by internalization into vesicles, lysosomal degradation and decrease of receptor mRNAs [65]. The GPCR cardiac β-adrenoreceptors became downregulated after prolonged *in vivo* infusion of catecholamines in rat [66]. GPCR α1-adrenoreceptors could be downregulated by the specific α1 agonist, R-(-)-N6-(2-phenylisopropyl)adenosine in rat atria inducing their uncoupling from G proteins and loss of $G_i$ proteins [67]. Desensitisation of rat heart contractility after sustained adenosine treatment seems to be mediated by the α1-adrenoreceptor and protein kinase C [68].

Continuous exposure of rat pancreatic islets to high glucose (300 mg/dl) induced glucose hypersensitivity after 3 hours which turned to glucose insensitivity after 6 hours of exposure [69]. Insulin receptor auto-antibodies (as agents able to provoke a sustained activation) induced an insulin-resistant state of glucose metabolism in 3T3-L1 adipocyte-like fatty fibroblasts after 6 hours of exposure blocking an early step in insulin signalling (but leaving insulin binding ability constant) [70].

Signalling of human cells is not more complex than that of e.g. *Caenorhabditis elegans* or *Drosophila melanogaster* because of more human signalling pathways, but because of much more cross-talks between signalling pathways in humans [71]. Due to this complexity desensitisation may often act on different pathways than that of the provoking agent. Signalling pathways often act as 'Darwinian competitors'. If one of them becomes stronger (e.g. by a cellular learning process), it induces molecular events (such as protein phosphorylation), which desensitise (inhibit, down-regulate, etc.) of 'competing' pathways. An example for this from the many is the heterologous desensitisation of G-protein coupled receptor (GPCR) and insulin-like growth factor pathways by insulin [72]. Conversely, chronic endothelin exposure desensitises the insulin pathway [73].

**Stress-induced desensitisation of signalling**

Desensitisation of a wide range of signals is occurring, if the cell or the animal experiences stress, such as heat shock, UV light, immobilization, or endoplasmic reticulum stress. Immobilization reduced the number of α1- and β-adrenoreceptors in rat hearts [74]. In agreement with the internalization → degradation sequence, immobilization stress first reduced the number of surface β-adrenoreceptors, and only then the total number of receptors [75]. Suppression of microRNA-16 gave a protection against acute myocardial infarction reversing β2-adrenergic receptor downregulation in rats [76]. Epidermal growth factor (EGF) receptor down-regulation was observed in the colon cancer cell lines SW480, HT29, and DLD-1 after ultraviolet light-C treatment inhibiting cell proliferation and survival [77]. UV light induced EGF, tumour necrosis factor (TNF) and interleukin-1 receptor down-regulation in mammalian cells activating the Jun-kinase cascade [78].

Insulin signalling desensitisation potentially leads to diabetes. Insulin receptor tyrosine phosphorylation was reduced by tunicamycin-provoked endoplasmic reticulum stress, which was reversed by the overexpression of activating transcription factor 6 (ATF6), a key signal of endoplasmic reticulum stress [79]. Similar, autophagy (but not proteasome) dependent down regulation of insulin signalling was observed after endoplasmic reticulum stress in fat tissue of obese human subjects and 3T3-L1 adipocytes [80]. Heat stress downregulated insulin signalling in pig testicular cells [81].

**Aging-induced signalling desensitisation**

Aging can be perceived as a cumulative result of the continuous stress by free radicals and other harmful effects inducing inflammation [82–86]. Aging is downregulating the renin-angiotensin system in rat kidneys [87]. Klotho-induced activation of the retinoic acid-inducible gene I/nuclear factor-κB (RIG-I/NF-κB) signalling pathway, as well as the subsequent production of proinflammatory mediators (tumour necrosis factor α and interleukin-6) and inducible nitric oxide synthase were reduced in the kidneys of aged senescence-accelerated mouse prone-8 (SAMP8) mice [88]. Downregulation of angiogenesis-related vascular endothelial growth factor (VEGF) signalling was reported in hearts of aging rats living a sedentary lifestyle, but was recovered in aging rats with exercise training [89]. Endoplasmic reticulum stress was activated in fatty livers of old mice by inhibiting hepatocyte nuclear factor 1 alpha (HNF1α) and downregulating farnesoid X receptor (FXR) [90]. Desensitisation of insulin receptor growth factor (in particular: downregulation of Irs1 and upregulation of Let-7 microRNA expression) was shown as a hallmark of the aged phenotype in developing B lymphocytes by a genome organization and chromatin study. These changes were associated with specific alterations in histone H3K27me3 occupancy, suggesting that Polycomb-mediated repression plays a role in precursor B cell aging [91].

**Learning, desensitisation, stress and aging in system-level signalling: phases of the same response?**

There are only a few reports of time-dependent changes in cellular signalling upon shorter *versus* longer extracellular signals. One of these was made on rat pancreatic islets, where 300 mg/dl, high concentrations of extracellular glucose induced a stronger response after 3 hours, which turned to glucose insensitivity after 6 hours [69]. This is clearly a two-step cellular response, where the pancreatic beta-cells first learned the presence of glucose and made a preconditioned, stronger response to them. However, after a longer time, an overload occurred and the cells turned to insensitive to glucose. Note that under natural conditions high glucose is only a temporary, postprandial event. A similar effect was observed, when 3T3-L1 cultured fat cells were exposed to insulin receptor auto-antibodies. Acute administration of anti-receptor antibodies induced a more efficient deoxyglucose uptake, while prolonged exposure led to insulin insensitivity [92]. Here again, high levels of the original agonist, insulin are also only transient, postprandial events.

If we take the examples of 1.) cellular learning and development of cellular memory after a few repeated stimuli [10,25,54,55]; 2.) the desensitisation of signalling after a prolonged exposure to the signal [61–70] and 3.) the result of the above studies [69,92] (where in an extended timescale first learning and then desensitisation was observed) together, the conclusion can be drawn that, in fact, cellular learning and desensitisation may be consequent phases of the same response. The cell first becomes more 'alert' and more 'ready' to respond to environmental changes. However, after a prolonged stimulus its signalling network becomes 'saturated' and starts to 'protect itself' (Fig. 2). We may also add stress [74–81] and aging [82–91] to this spectrum, where 'fatigue' of the signalling network is induced by both as examples of overloading short term (stress) and long-term (aging) changes (Fig. 2). We note that comparative studies of agonist-, stress- and aging-induced desensitisation are missing. Therefore their combination on Fig. 2 is only illustrative and hypothetical.

**Discrimination between repeated and unexpected signals: perhaps as also a property of single cells?**

Desensitisation protects the system from the overload of inputs. At a low level of complexity overload can be understood that too many of the same signal within a certain time (where the system may adjust its thresholds defining the "too many" and the "within a certain time"). At a higher level of complexity overload also occurs, if the system is not able to make 'groups' of similar input patterns. In fact, our brain defines objects (features, categories, concepts, etc.) as groups of correlating 'suspicious coincidences'. Moreover, recognition of (and reduced response to) similar input patterns helps to highlight unexpected signals, which is essential for survival. If a layer of Hebbian learning units becomes connected by modifiable anti-Hebbian feed-backs, the resulting system is able to learn this discrimination and to recognize other principal components of an incoming, complex signal than only its first principal component [93,94]. A well-known biological example is that of the mormyrid electric fish, which is able to eliminate predictable inputs produced by its own, regular motor output. However, this response is a general feature of cerebellum-like, laminar structures, where anti-Hebbian outputs of a deeper layer modulate outer layers (Fig. 3A) [95]. Thus using anti-Hebbian learning prevents excessive noise (i.e. regular, correlating, expected input) from masking important (i.e. unexpected) sensory information. Most sensory systems work based on the principle of fold-change detection, which allows for a proportional response to the fold-change of a signal (the unexpected) relative to the background (the repeated) [96]. From the complexity of learning responses of non-neuronal single cells [2] and the presence of distributed decision making in cellular signalling [97], we may expect that the widespread occurrence of anti-Hebbian learning in signalling networks (see examples above) is involved in the discrimination between repeated and unexpected signals in single, non-neuronal cells, too. Horizontal activation at a receptor-proximal level, as well as mutual inhibition at a receptor-distant level in signalling networks also point toward this expectation. For instance, in the TNF-induced NF-κB signalling, the well-studied upstream crosstalk conveyed by TNFR-associated factors (TRAFs) acts as horizontal activation at the receptor proximal level [98]. While downstream, a network motif containing inhibition has been described that can impart fold-change detection to cell signalling circuits: the incoherent type-1 feedforward loop (I1-FFL) (Fig. 3B) [96, 99, 100]. I1-FFL is one of the most frequently occurring network motifs in transcriptional networks [101]. Besides fold-change detection, I1-FFLs have a role in response acceleration even in yeast [102]. In an I1-FFL, X upregulates Y, while it also upregulates Z, a repressor of Y. This indirect repression of Y, coupled with the direct activation of Y, can be considered an anti-Hebbian learning mechanism. Besides NF-$\kappa$B signalling, I1-FFLs were also suggested to enable fold-change detection in the nuclear levels of the transcription factors of transforming growth factor beta (TGF-$\beta$) signalling, explaining how the cells are able to give the same proportional response, even though the nuclear level of transcription factors can vary greatly from cell to cell [96]. The occurrence of I1-FFLs in major signalling pathways suggests that this learning mechanism may be a rather general feature of signalling networks. Even still, to decide whether discrimination between repeated and unexpected signals is also a property of single cells, future experiments are required.

**Applications of cellular learning and 'forgetting' in pharmacology and drug design**

Mimicking cellular learning (memory) became a recent hit in drug design. Chemically induced proximity between two adjacent signalling proteins (a new drug design paradigm [103,104]) is actually copying Hebbian learning of the cell [10]. In this 'cellular learning scenario' chemical proximity-developing drugs induce targeted posttranslational modifications of key, otherwise undruggable proteins. In the reversed, anti-Hebbian learning model, chemically induced proximity promotes the selective degradation of the target [105,106].

Drug resistance can be conceptualised in a learning network model as habituation. Biological networks may contain nodes, where stimulation breaks the habituation (drug resistance) developed by the network [52]. Limited drug tolerance can be conceptualised as sensitisation in a learning network model, most simply by displaying a hysteresis-type response. Interestingly, breaking of sensitisation was much rarer phenomenon in a model of 35 biological networks than that of habituation [52]. However, the break of allergy-induced sensitisation against drugs became a carefully manageable clinical modality in the last decades – as we will describe in the following paragraphs.

Signal desensitisation plays a major role in anti-cancer therapy, which can be regarded as 'the archetype' of therapeutic intervention consequences in a number of other diseases. As an example for the first modality of protocols, the anti-cancer agent, 90 kDa heat shock protein (Hsp90) inhibitors induce a desensitisation of the EGF receptor *via* p38 MAPK mediated phosphorylation at Ser1046/1047 of the EGF receptor in human pancreatic cancer cells. Here drug-induced desensitisation of the cancer-promoting growth factor signal is a mode of action to avoid disease [105]. As an example for the second modality of consequences, gastric cancer cells become desensitised to trastuzumab-treatment by upregulation of MUC4 expression and by catecholamine-induced β2-adrenoreceptor activation. Here desensitisation, i.e. the development of drug resistance is an unwanted consequence of drug treatment [106]. As a third modality of therapeutic interventions, response-desensitisation (i.e. breaking the sensitisation of unwanted side-reactions for the drug) is a general goal in cancer therapy, where patients often develop sensitivity towards the administered drugs [107,108].

Supporting the notion that cancer therapy experiences are '*pars pro toto*' for other conditions, increased hypersensitivity to drugs (e.g. for aspirin and non-steroid anti-inflammatory drugs,

NSAIDs in patients with heart disease or inflammatory diseases; for insulins, penicillin or other antibiotics) became a general phenomenon in the past 25 years due to the widespread and intensive drug use. Therefore, carefully administered drug-desensitisation protocols became more and more important in the clinical practice [109,110]. Similarly, drug induced desensitisation of cellular mechanisms of action is also a general phenomenon in a number of non-cancer treatment protocols, including that of β-adrenergic agonists in asthma [111], diabetes [112,113] or heart failure [114].

**Conclusions**

In our previous work [10] we gave several examples for cellular learning [3–9] (i.e. the formation of cellular memory [25,52–55]), and showed that cellular learning can be perceived as Hebbian learning [48] of signalling networks, where learning is accompanied by strengthening of those protein-protein, protein-microRNA and chromatin interactions, which participate in the learning process [10]. In this review we summarize our current knowledge on the other side of the coin: when signalling networks 'refuse' to learn more, and become desensitised by prolonged presence of the provoking signal, by stress or aging.

Cellular learning proceeds using several molecular mechanisms, such as conformational memory of intrinsically disordered proteins (IDPs), prions and mnemons [10–16], protein translocation [18–21], as well as miRNAs and lncRNAs [6,7,22–24]. System level responses of cellular learning include chromatin memory [3,5,8,9,25–32], signalling protein kinase cascades [33–38], as well as network responses other than those of signalling networks, such as metabolic reaction networks and metabolons [40–47].

Besides Hebbian learning [10] by the molecular mechanisms listed in the previous paragraph, non-neural cells also display anti-Hebbian learning, where connection strengths decrease between the signalling network components [23,31,51,52]. Cells are also able to 'forget' using TET DNA demethylases, lncRNAs or molecular chaperones [58–60] besides developing a cellular memory.

A cellular habituation-type of response is the development of desensitisation. This often involves first decoupling of signalling network components from each other followed by receptor internalization and downregulation [61–70]. This may be displayed by cross-desensitisation, where prolonged exposure for a pathway agonist induces the desensitisation of other pathway(s) [72,73]. A specific condition of generally increased signal intensity and/or complexity is stress, which desensitises a wide range of signals [74–81]. Aging can be perceived as a result of cumulative stress [82–86], which downregulates a large number of signalling pathways [87–91].

Here we propose that cellular learning, desensitisation, stress and aging may be placed as responses along the same axis of more and more intensive (more and more prolonged, or more and more often repeated) signals (Fig. 2). We pose the question, whether single cells may also display discrimination between repeated and unexpected signals, a common property of neuronal and artificial neural networks (Fig. 3A) [93–95]. As a first step in answering this question, we present fold-change detection enabling I1-FFLs as anti-Hebbian learning mechanisms that are potentially general features of signalling networks given their occurrence in prominent signalling pathways like NF-*κ*B (Fig. 3B) and TGF-*β* signalling [96, 98-101].

Finally, we summarize applications of signalling network learning and desensitisation in clinical treatments discriminating between five scenarios (Fig. 4): 1.) when cellular Hebbian learning is mimicked by chemically induced proximity between signalling network components [103,104]; 2.) when cellular anti-Hebbian learning is mimicked by chemically induced proximity of protein degradation [104,105]; 3.) when desensitisation of unwanted signalling (such as that in cancer) is the mechanism of drug action [105]; 4.) when desensitisation of wanted signalling occurs, and should be avoided (in cancer, asthma, diabetes or heart failure [106, 110–114]); and finally, 5.) when sensitisation against a drug occurs by allergic reaction, which also should be minimized (in cancer, inflammatory diseases, diabetes or infections [107,109,110]).

We hope that our summary will prompt further investigations of the phenomena, when cells learn (develop cellular memory) by Hebbian learning type processes, and when they 'refuse' to learn more, i.e. become desensitised (display anti-Hebbian learning, i.e. cellular 'forgetting') by prolonged exposure to environmental signals, by stress or by aging. It is an interesting question, how much desensitisation remains specific for the given pathway, and how much it is displayed as cross-desensitisation of other pathways, or as a general forgetting (desensitisation) of many (if not all) pathways. While agonist-induced desensitisation is mostly the former, directed type desensitisation against the same pathway (or selected different pathways), stress- and aging-induced desensitisation are usually more widespread phenomena involving a larger segment of the signalling network. We predict that network methodologies will greatly help the discrimination between these scenarios.

**Declarations**
**Author contributions** Peter Csermely initiated the idea, wrote the initial draft and finalized the manuscript. Tamás Veres and Márk Kerestély finalized paper figures and contributed to key concepts. All authors participated in the interpretation of initial ideas and writing the manuscript.
**Conflict of interest** The authors declare no conflict of interest.
**Ethics approval and consent to participate** This study required no ethics approval and consent to participate.
**Consent for publication**. All authors provided consent for publication of this work.
**Availability of data and material**. This review study does not contain parts which require the deposit of data or other material.
**Funding**. This work was supported by a grant from the Hungarian Science Foundation (OTKA K131458) and by the Thematic Excellence Programme (Tématerületi Kiválósági Program TKP2021-EGA-24) of the Ministry for Innovation and Technology in Hungary, within the framework of the Molecular Biology thematic program of the Semmelweis University.

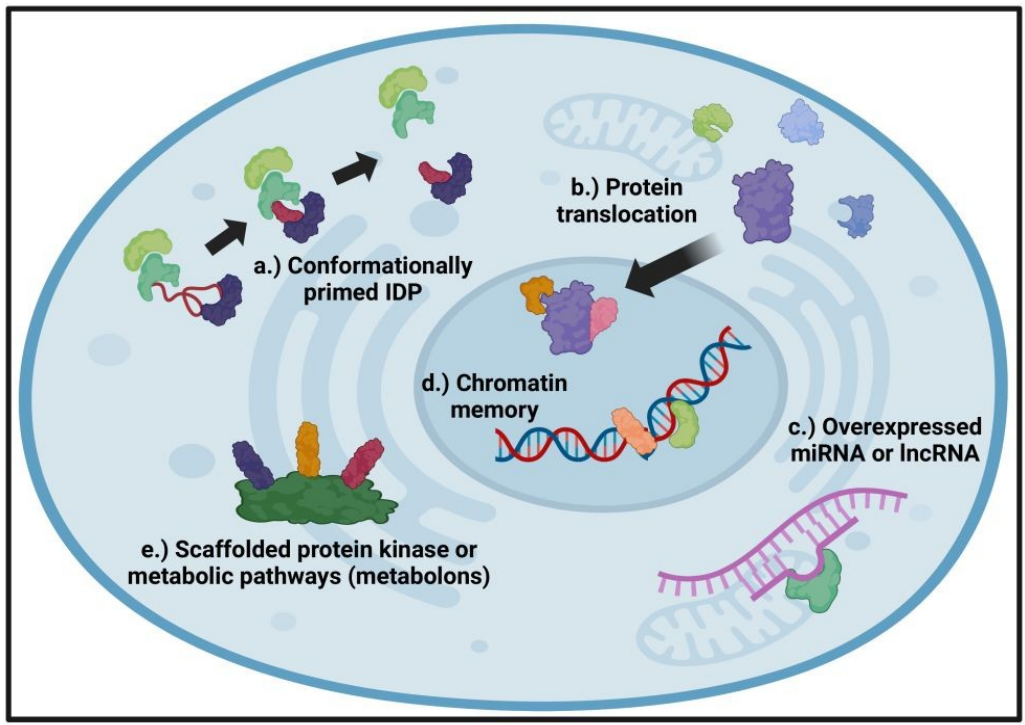

**Fig. 1** Molecular mechanisms of connection strength increase by cellular Hebbian learning-type processes. a.) Stronger and faster binding of 'conformationally primed' IDPs (prions and mnemons) to their signalling partners. b.) New connection sets of translocating proteins. c.) New connections of overexpressed miRNAs and lncRNAs. d.) Chromatin memory. e.) Scaffolded protein kinase and metabolic pathways. This figure was created with BioRender.com.

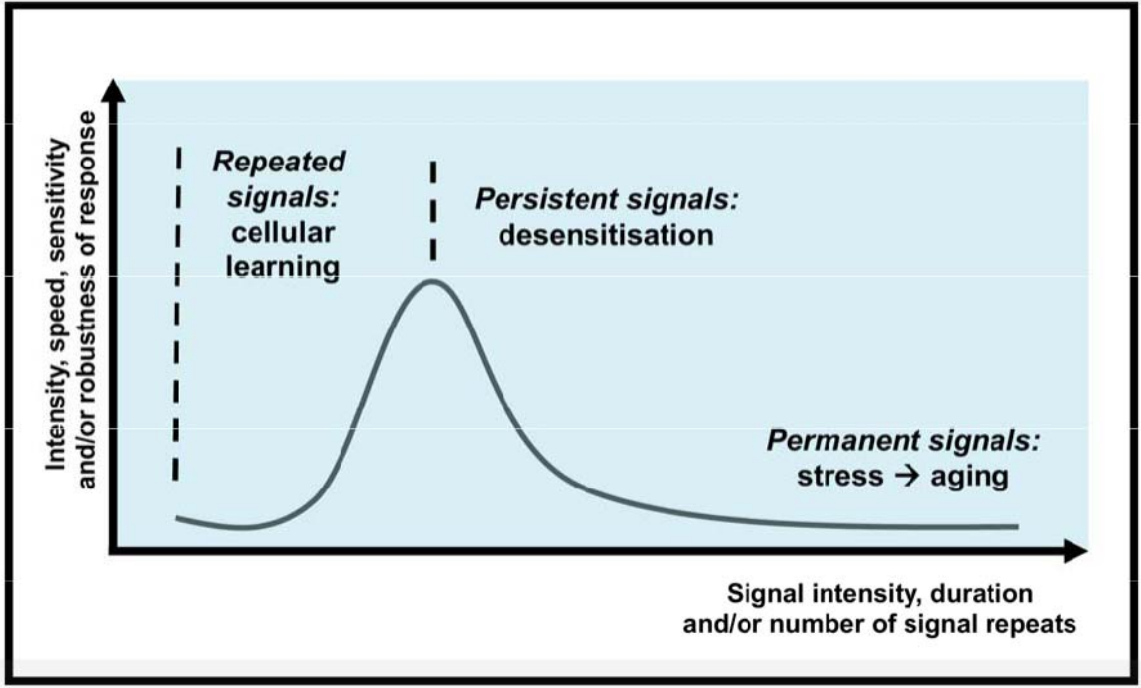

**Fig. 2** Repeated signals induce cellular learning; persistent signals lead to cellular desensitisation; permanent signals (such as the accumulated signals and damage in aging) overload the signalling network and provoke cellular stress. Note that on the contrary to the few studies showing a change from cellular learning to desensitisation in the same system [69,92], comparative studies of agonist-, stress- and aging-induced desensitisation are missing. Therefore, this figure is only hypothetical, illustrative and by no means quantitative.

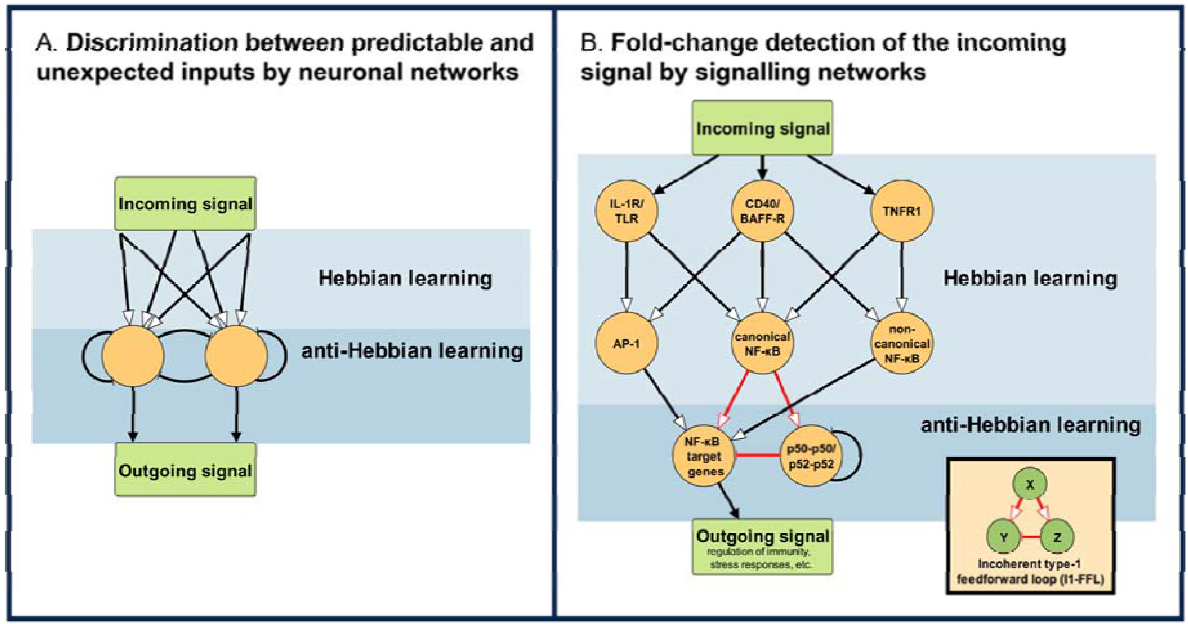

**Fig. 3** Hebbian and anti-Hebbian learning layers in neuronal and signalling networks. A) Schematic representation of the combination of Hebbian- and anti-Hebbian learning layers, which result in the discrimination between predictable and unexpected inputs. The cerebellum-like laminar structure of the figure is widespread in various animal and human neuronal networks [95] and was also shown to work in computational neural networks [93,94]. Note that self-inhibitory connections ('autapses') are not necessarily needed for the circuit. B) Proposed Hebbian and anti-Hebbian learning layers in the NF-$\kappa$B signalling. Receptor proximally, the signalling of multiple receptors can lead to the activation of the NF-$\kappa$B pathway through TRAFs. Concurrently, some can lead to the activation of the adjacent AP-1 signalling pathway [98]. This constitutes a horizontal activation in the proposed Hebbian learning layer in the upstream signalling. In the downstream signalling, AP1 and the non-canonical NF-$\kappa$B pathway modulate NF-$\kappa$B target genes (e.g., interleukin-8, interleukin-6). The canonical NF-$\kappa$B (RelA-p52 heterodimer) has also been shown [100] to upregulate the formation of the transcriptionally inactive p50-p50/p52-p52 homodimers that act as competitors to NF-$\kappa$B for $\kappa$B sites in the target genes' promoters. These interactions, highlighted in red, constitute a Type-1 incoherent feed-forward loop (I1-FFL, see inset) that can be understood as an anti-Hebbian learning mechanism. This system enables fold-change detection of the incoming signal in NF-$\kappa$B nuclear levels, that is analogous to discrimination between predictable and unexpected inputs. This figure was created with yEd Graph Editor.

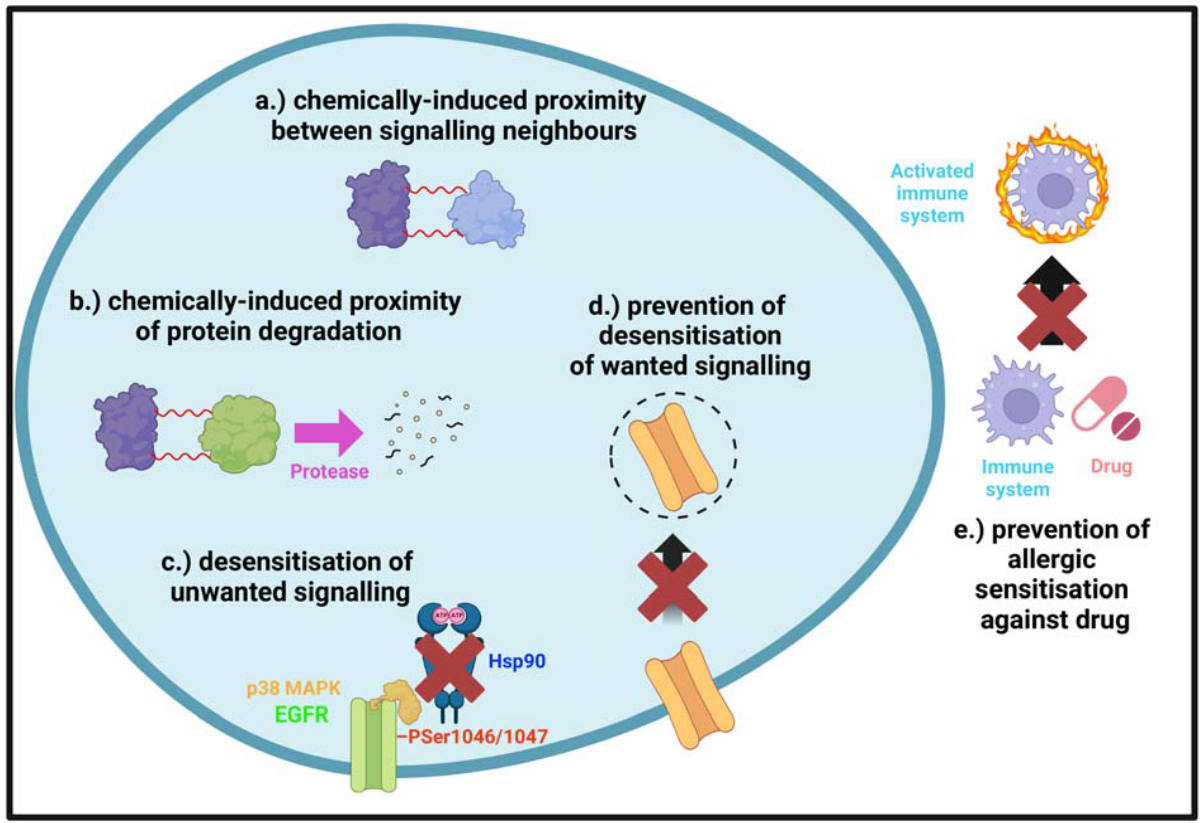

**Fig. 4** Clinical treatments for and against Hebbian and anti-Hebbian learning of the signalling network. a.) Cellular Hebbian learning is mimicked by chemically induced proximity between signalling network components. b.) Anti-Hebbian learning is mimicked by chemically induced proximity of protein degradation. c.) Desensitisation (drug-induced anti-Hebbian learning) of unwanted signalling (such as that in cancer). d.) Prevention of desensitisation of wanted signalling (cellular anti-Hebbian learning) in cancer, asthma, diabetes or heart failure. e.) Prevention of allergy-induced sensitisation (Hebbian-learning) against the drug in cancer, inflammatory diseases, diabetes or infections. This figure was created with BioRender.com